\documentclass[a4paper,11pt] {article}
\usepackage{graphicx}
\usepackage{authblk}
\usepackage{xcolor}
\usepackage{caption}
\usepackage{pgfplots}
\pgfplotsset{compat=1.3}
\usepackage{hyperref}
\usepackage{url}
\usepackage{siunitx}
\usepackage{tabularx}
\usepackage{commath}
\usepackage{mathtools}
\usepackage{color}
\usepackage{listings}
\usepackage{wrapfig}
\usepackage{natbib}
\usetikzlibrary{plotmarks}
\usepackage{wrapfig}
\usepackage{multicol}
\usepackage{afterpage} 
\usepackage{natbib}
\usepackage{lineno}

\usepackage{graphicx}
\usepackage{hyperref}

\usepackage{amsmath}
\usepackage{array} 
\usepackage{aas_macros}

\makeatletter

\newenvironment{figurehere}
  {\def\@captype{figure}}
  {}
\makeatother
\title{\textbf{Cross validation of albedo determination for 1627 Ivar from three different techniques}}

\author[1]{Elena Selmi}
\author[2]{M. Devog\`{e}le}
\author[3]{J. R. Masiero}
\author[4]{N. Vega Santiago}
\author[5]{E. L. Wright}
\author[6]{M. Ferrais}
\author[6]{E. Fernández-Valenzuela}
\author[7,8]{G. Borisov}
\author[9]{Ph. Bendjoya}

\author[9]{J.-P. Rivet}
\author[9]{L. Abe}
\author[9]{D. Vernet}
\author[10]{A. Cellino}

\affil[1]{University of Oxford, Department of Physics, Parks Rd, OX1 3PU, Oxford, UK; elena.selmi@lmh.ox.ac.uk}
\affil[2]{ESA NEO Coordination Centre, Largo Galileo Galilei, 1, 00044 Frascati (RM), Italy}
\affil[3]{Caltech/IPAC, 1200 E California Blvd, MC 100-22, Pasadena, CA 91125, USA}
\affil[4]{University of Arizona, Planetary Science Department, 1629 E University Blvd, Tucson, AZ 85721}
\affil[5]{UCLA Astronomy, PO Box 951547, Los Angeles CA 90095-1547}
\affil[6]{Florida Space Institute (UCF), 12354 Research Parkway, Orlando, FL 32826 (USA)}
\affil[7]{Institute of Astronomy and National Astronomical Observatory, Bulgarian Academy of Sciences, 72, Tsarigradsko Chauss\`{e}e Blvd., Sofia BG-1784, Bulgaria}
\affil[8]{Armagh Observatory and Planetarium, College Hill, Armagh BT61 9DG, Northern Ireland, UK}
\affil[9]{Université Côte d’Azur, CNRS, OCA, LAGRANGE, France}
\affil[10]{INAF, Osservatorio Astrofisico di Torino, via Osservatorio 20, 10025 Pino Torinese, Italy}

\date{}

\begin{document}

\maketitle
\vspace{-1.5cm}
\begin{abstract}
Near Earth Asteroids are of great interest to the scientific community due to their proximity to Earth, making them both potential hazards and possible targets for future missions, as they are relatively easy to reach by
spacecraft. A number of techniques and models can be used to constrain their physical parameters and build a comprehensive assessment of these objects. In this work, we compare physical property results obtained from improved $H_V$ absolute magnitude values, thermophysical modeling, and polarimetry data for the well-known Amor-class NEO 1627 Ivar. We show that our fits for albedo are consistent with each other, thus demonstrating the validity of this cross-referencing approach, and propose a value for Ivar's albedo of $0.24^{+0.04}_{-0.02}$ . Future observations will extend this work to a larger sample size, increasing the reliability of polarimetry for rapid asteroid property characterization, as a technique independent of previously established methods and requiring significantly fewer observations.
\end{abstract}
\begin{multicols}{2}
   \section{Introduction} 

   Near-Earth Objects (NEOs), which Near-Earth asteroids make up the most part of, are objects with perihelia (q) $\le$1.3 astronomical units (au), bringing them close to Earth.  This proximity has gained them wide-spread attention from the scientific community \citep[for a review, see][]{binzel15} as it allows for detailed studies of their physical properties and probing size regimes much smaller than can be studied for more distant populations. The NEOs are a dynamical, evolving population, which by definition can potentially impact Earth, while at the same time being reachable by spacecraft as demonstrated by many missions, such as OSIRIS-REx \citep{lauretta17}, Hayabusa2 \citep{watanabe17} and DART \citep{rivkin21}. Their proximity also allows for a variety of observational techniques to be applied in studying them, through which their characteristics can be accurately modelled. These parameters include shape \citep[e.g.][]{Magri_2011,Kueny_2023}, diameter \citep[e.g.][]{Trilling_2008}, period \citep[e.g.][]{Thirouin_2016}, albedo \citep[e.g.][]{Masiero_2021}, and thermal inertia \citep[e.g.][]{Fenucci_2023}, and are fundamental to build an assessment of both the hazard these objects may represent \citep[e.g.][]{Dotson_2024} and the potential for the foreseeable in-space resource utilization \citep[e.g.][]{Xie_2021}.

\begin{figure*}[b!]
  \centering
  \includegraphics[width=0.45\linewidth, height = 0.39\linewidth]{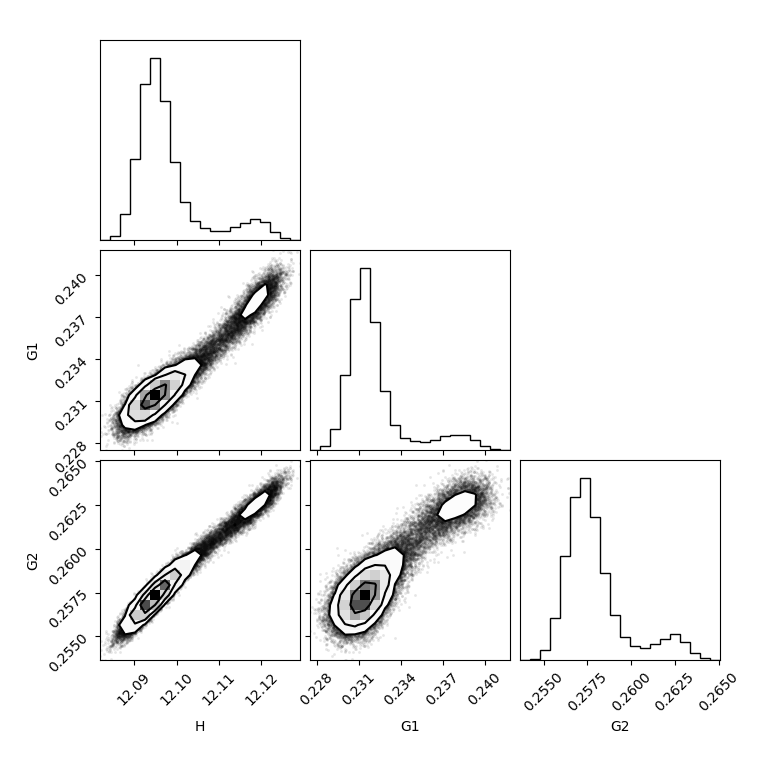} \quad
  \includegraphics[width=0.49\linewidth, height = 0.39\linewidth]{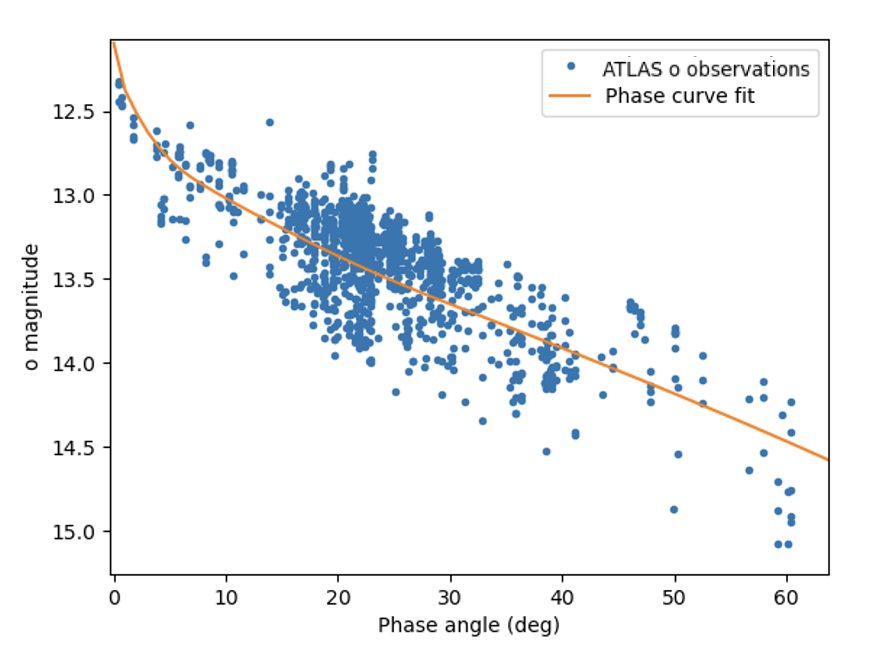} \\
  
  \caption{Corner plots and progression of the MCMC routine fitting for $H$, $G_1$ and $G_2$ (left) and fitted phase curve of Ivar (right), with the data extracted from the SSCAT.}
  \label{fig:fig1}
\end{figure*}

In this work, we use three different techniques to assess some of the physical properties of Ivar (1627) (Section\ref{sec.ivar_intro}): refined $H$ magnitude values and thermophysical modeling of multi-epoch infrared data to simultaneously constrain the size, albedo, and thermal inertia \citep{wright07, koren15}, and polarimetry, which allows us to directly constrain the albedo of the surface (\citet{cellino15} and \citet{masiero12}).  Thermophysical modeling requires thermal infrared detections over multiple viewing geometries to constrain diameter and thermal inertia, while precision absolute magnitude determination requires knowledge of an object's rotational light curve and coverage of its photometric phase curve across a wide range of angles, including near opposition.  In contrast, a single high-phase-angle ($\alpha>30^\circ$) polarimetric measurement of an asteroid can be sufficient to provide a robust measurement of the surface geometric albedo. The latter provides a constraint on compositional taxonomy, as per \citet{mainzer11tax}, and when combined with a highly precise absolute magnitude it also allows for estimation of the object's diameter.  We discuss the data required and our comparison of these techniques below.

    \subsection{1627 Ivar}
    \label{sec.ivar_intro}
    Ivar (1627) is a Near Earth asteroid that was discovered in 1929 by E. Hertzsprung at Johannesburg (as per the Minor Planet Centre\footnote{\url{https://www.minorplanetcenter.net/db_search/show_object?object_id=1627}}). It belongs to the Amor class of NEOs (q$>$1.017 au but $<$1.3 au, thus at all times orbiting beyond Earth) and the taxonomic class Sqw \citep{jones18}. Its synodic rotation period was found to be 4.8 hours by \citet{harris85}, but is now accepted by the literature to be 4.795 h \citep[as found by][]{warner15}. More recently, shape modeling of Ivar using a large set of lightcurves provided a sidereal period of P=4.79517 hours \citep[][as per \citet{durech10}\footnote{\url{https://astro.troja.mff.cuni.cz/projects/damit/}}]{kaasalainen04,hanus15}.  It is the first asteroid whose size was constrained by radar echos \citep{ostro90}, which found it to be about twice as long as it is wide, with a maximum axis of minimum 7 km and probably within $20\%$ of 12 km. Models approximating to a spherical shape have found an effective diameter consistent with this value, with \citet{mainzer14} setting it at about 8.485 km and \citet{hanus15} at 7.7 $\pm$0.6 km. The radar- and lightcurve-based model presented by \citet{crowell17} allows for an elongated shape and found the maximum extensions along the three body fixed coordinates to be (15.15 × 6.25 × 5.66 km) ± 10$\%$ (volumetric spherical equivalent diameter\footnote{from $\frac{4\pi}{3}r^3 = \frac{4\pi}{3}abc \Rightarrow D = \sqrt[3]{ABC}$, where $a$, $b$ and $c$ are the semi-major axes and $A$, $B$ and $C$ the maximum elongations.\label{D_formula}} = $8.12\pm0.47$ km). Ivar is a widely studied object, making it an ideal candidate for validation of our cross-reference approach, as it can be tested against a range of previous measurements.

    \section{Photometric measurements}
    \label{sec.hmag}

We extracted observations of Ivar from the version two of the Asteroid Terrestrial-impact Last Alert System (ATLAS) \citet{Tonry_2018} Solar System Catalog (SSCAT). The SSCAT contains observations obtained by all the ATLAS telescopes and are obtained in either the o (orange) or c (cyan) bands. We fitted these observations using the $H$, $G1$, $G2$ model of the asteroid phase function \citep[equations 18 and 19 of][]{muinonen10} using a Markov Chain Monte Carlo (MCMC) routine following the \citet{Vega_2023} method. This method consists of using prior information on the expected $G_1$ and $G_2$ parameters based on the results from \citet{Mahlke_2021}.

In \citet{Mahlke_2021} the $G_1$ and $G_2$ parameters for more than 100 000 asteroids were obtained using ATLAS data. As Ivar is an S-type asteroids, we extracted the $G_1$ and $G_2$ distributions for all the S-type in \citet{Mahlke_2021} and used it as a prior information for the expected $G1$ and $G2$ for Ivar. The distributions of $G_1$ and $G_2$ for the S-types and other taxonomic types can be viewed on Fig. 6 of \citet{Mahlke_2021} paper. No prior is used on H. The result of the MCMC fit using prior information for Ivar provides an $H$ magnitude (in the o band) of $12.10 \pm 0.02$, $G_1 = 0.23 \pm 0.01$, and $G_2 = 0.26 \pm 0.01$. Fig. \ref{fig:fig1} represents the progression of the MCMC routine on the left and, on the right, reports Ivar photometric measurements in the o band extracted from the SSCAT with our best fit of the $H$, $G_1$, $G_2$ model to the data. 
    
To obtain the $H$ magnitude in the V band, we need to correct for the color differences in between these two bands. From the Minor Planet Centre\footnote{\url{https://minorplanetcenter.net/iau/info/BandConversion.txt}}, the colors' correction between the V and o bands is 0.33 magnitude. In a recent effort to improve color corrections for photometric observation obtained in different band, \citet{Hoffmann_2024} found the same correction of 0.33 mag between the o and the V band. Our determination of the $H$ magnitude of Ivar is thus $H_V = 12.43\pm0.02$. This value is 0.40 magnitude brighter than the $H$ magnitude reported on the Minor Planet Center database, which is 12.83. Other previous works report it to be 12.83 (Small-Body Database Lookup\footnote{\label{Small_footnote}\url{https://ssd.jpl.nasa.gov/tools/sbdb\_lookup.html}}), $12.87\pm0.10$ \citep{mueller11} or 13.00 \citep{carry16}. This difference in $H$ magnitude can be due to multiple reasons, as seen in \citet{mahlke21}. In particular, it is relevant to note how the viewing geometry can affect the result, as well as fixing the parameters of the phase function (with regards to the latter, note that the MPC is currently using the HG model, not the $H$, $G1$, $G2$ one). 

\begin{figurehere}
    \centering
    \includegraphics[width = \linewidth]{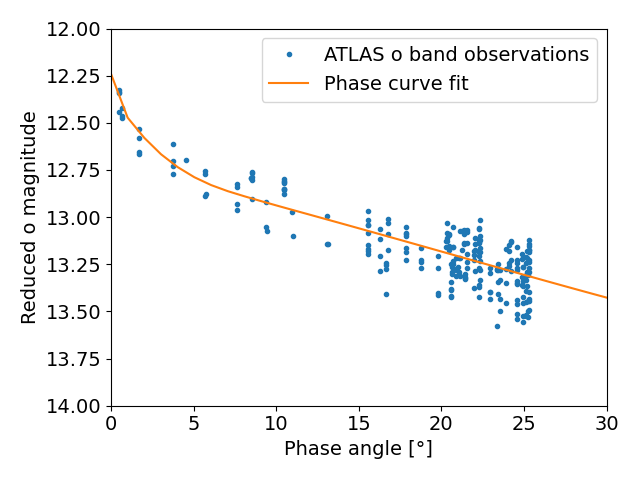}
    \caption{Fitted phase curve using only the best ATLAS apparition.}
    \label{fig:atlas}
    \end{figurehere}
    
We have probed how this affects our own result in two ways. Limiting ourselves to the best apparition in the ATLAS data (hence decreasing the scattering, see Fig. \ref{fig:atlas}), we found that our MCMC routine gives $H_V = 12.57$. We then applied the model from \citet{carry24}, that uses a new parameter to take into account the viewing geometry and fits all the data at once from multiple apparitions, to both the ATLAS and the ZTF \citep{masci18} data. We obtain $H_V = 12.57$ from ATLAS and $H_V = 12.63$ and $H_V = 12.64$ from ZTF's g and r band respectively. We thus note that results for Ivar's absolute magnitude vary significantly based on datasets and apparition and should be considered carefully when using them to infer NEOs' albedos.

Using the volume-based spherical equivalent diameter from Section \ref{sec.ivar_intro} and $H_V = 12.43\pm0.01$, we obtain an albedo for Ivar of $p_{\rm{V}} = 0.29\pm0.03$. The same diameter applied to the MPC value gives $p_{\rm{V}} = 0.20\pm0.02$. Further discussion of this range of values can be found in Section \ref{sec.tpm} and \ref{sec.comparison}.

   \section{Thermophysical Modeling}
   We applied a thermophysical model to observations of Ivar made by the Near-Earth Object WISE (NEOWISE) mission and present here the results.
   \subsection{The model}
      \label{sec.tpm}

   Thermophysical models (TPM) have become predominantly used in more recent years, with infrared data becoming available for multiple NEAs. They differ from Standard Thermal Models for the free-parameterization of some of STM's assumptions such as the spherical surface shape, as well as for the addition of new parameters, like the thermal inertia. They neglect lateral thermal diffusion, meaning that a 1-D diffusion equation can be solved numerically until specified convergence criteria are met; the best-fitting parameters are determined by considering the minimized sum of square deviations, $\chi^2$ (for greater detail, see \citet{delbo15}). 
        
   Although TPMs generally constrain NEO parameters to a better degree than simple STMs, it is important to underline that because of the larger number of fittable parameters and the interaction between them, they do so only in cases where a large amount of infrared data is available for an object. If such a dataset isn't available, simple thermal models are usually preferred, as they provide sufficient physical parameter constraints, but have significantly smaller computational requirements and fewer opportunities for bad fits due to under-constrained models \citep{masiero19}.  Here we use the rotating, cratered thermophysical model introduced by \citet{wright07}, which uses synthetic hemispherical unresolved craters to represent the surface roughness. Employing an affine-invariant Markov Chain Monte Carlo simulation \citep[adapted from][]{foreman13}, this TPM can produce best fits for up to 10 different free parameters — the longitude and latitude of the spin axis pole position, diameter, geometric albedo, rotational period, thermal inertia, cratering fraction, $p_{IR}/p_V$ ratio, and b/a and c/b axis ratios. Note that in this study, the input rotational period of 4.795 hours was kept fixed, to reduce the number of free-parameters and thus the computational requirements, and it won't be further discussed. 

   Different priors are assumed for the parameters listed above. The one for  the diameter is a log-uniform distribution of values between 1 m and 1000 km, and the one for visual albedo is a mixture model of two Rayleigh distributions given by\citet{wright16}. The prior for the cratering fraction ($f_c$) is uniform in 0–1 and the one of the thermal inertia ($\Gamma$) is a log-uniform distribution for values of 2.5–2500 $J m^{-2} s^{-\frac{1}{2}} K^{-1}$ with a width of one unit of natural log. All other priors are described in \citet{satpathy22}.
   \begin{table*}[t!]
    \centering
    
    \begin{tabular}{|c|c|c|c|c|c|}
        \hline
        Date (yy-mm-dd) & \parbox[c]{2cm}{Number of \\ observations} & Number of  channels & \parbox[c]{2cm}{Average heliocentric \\ distance (AU)} & \parbox[c]{3cm}{Object-observer \\ distance (AU)} & \parbox[c]{2cm}{Solar phase \\ angle (deg)} \\
        \hline
       2010-05-07 & 13 & 4 & 2.1141	&1.3225 & 28.3651\\
        
        2014-01-20 & 14 & 2 & 1.9772 &1.2159 &	29.8107\\
        
        2014-11-26 & 7 & 2 &2.5869 & 1.7056 &	22.4143\\
        
        2015-04-24 & 21 & 2 & 2.2828& 1.4487 & 25.7368\\
        
        2016-10-10 & 10 & 2 & 2.2425 & 1.4188 & 26.4450\\
       
        2018-01-19 & 18 & 2 & 1.9940 & 1.2278 & 29.5683\\
        
        2018-07-08 & 23 & 2 & 1.1671 & 1.0069 & 54.7105\\
        
        2018-07-21 & 1 & 2 & 1.1390 & 1.0569 & 58.7870\\
        
        2018-08-20 & 1 & 2 & 1.1312 & 1.0495 & 58.6111\\
        
        2018-09-09 & 102 & 2 & 1.1705 & 0.9781 & 52.8537\\
        
        2018-11-01 & 36 & 2 & 1.3936 & 0.9265 &	41.6143\\
        
        2018-11-17 & 2 & 2 & 1.4781	& 0.9570 & 39.8666\\
       
        2019-11-15 & 13 & 2  & 2.6008 & 1.7140 & 21.8974\\
        
        2020-04-01 & 17 & 2 & 2.4382 & 1.5578 &	22.0993\\
       
        2021-10-03 & 13 & 2 & 2.0919 & 1.3046 & 28.5810\\
       
        2022-02-18 & 10 & 2 & 2.5093 & 1.6213 & 21.5424\\
       \hline

    \end{tabular}
    \caption{Epochs of observations remaining after the filtering process. These were used as input for the Thermophysical Model.}
    \label{tab:table1}
    \end{table*}

   \subsection{NEOWISE}
   Launched in 2009, the Wide-field Infrared Survey Explorer (WISE) was a mission funded by NASA's Astrophysics Division to map the sky in four infrared bandpasses with a 40-cm telescope \citep{wright10}. It scanned circles of the sky from low-Earth orbit, maintaining a viewing geometry near 90$^\circ$ Solar elongation, and precessing over a 1 year period with the orbit of the Earth; this pattern covers the whole sky in half a year as the Earth–Sun line turns by 180°. A low-Earth orbit was chosen to allow for transmission of large quantities of data to the ground while reducing mission costs.
    \begin{table*}[b!]
    \centering
    \begin{tabular}{|c|c|c|c|c|c|}
        \hline
        Model & Diameter (km) & Albedo & Theta ($Jm^{-2}s^{-1/2}K^{-1}$) &  Crater Fraction & $p_{IR}/p_V$ \\
        \hline
       Spherical &$7.914\pm0.418$ & $0.28\pm0.10$ & $4.594\pm2.407$ & $0.182\stackrel{+0.604}{\underset{-0.144}{}}$ & $1.715\stackrel{+0.304}{\underset{-0.266}{}}$\\
       \hline
        Triaxial & $6.700\pm0.935$ & $0.36\pm0.15$ & $3.743\pm2.785$  & $0.473\stackrel{+0.354}{\underset{-0.322}{}}$ & $1.424\stackrel{+0.389}{\underset{-0.289}{}}$\\
        \hline
        
    \end{tabular}
    \caption{Ivar's physical parameters obtained via Thermophysical Modeling for the spherical (up) and triaxial (down) model. Theta corresponds to Thermal Inertia. We ignore the results for the pole position, as this was poorly constrained.}
    \label{tab:table2}
    \end{table*}

   The survey was initially conducted at four effective wavelength bands (3.4, 4.6, 12 and 22 $\mu m$, known as W1, W2, W3 and W4 respectively), but after September 2010, when the cryogen chambers ensuring the operating temperature for the 2 longest wavelengths were exhausted, it was only continued at W1 and W2 only. After being placed in hibernation in 2011, the mission was reactivated by NASA's Planetary Science Division in 2013 and renamed NEOWISE \citep{mainzer14}, and its main mission became identifying and characterizing the population of Near-Earth Objects. Operations were stopped in 2024, with the current database containing 1.6 million measurements of 44,00 different solar system objects, including 1,598 NEOs (as per the mission website\footnote{\url{https://neowise.ipac.caltech.edu}}). Significantly, the long operational period in almost unchanging environmental conditions has resulted in a well-calibrated and uniform dataset, with many objects having multiple detections from a range of geometries and orbital locations.

    \begin{figure*}[t!]
  \centering
  \includegraphics[width=0.45\linewidth]{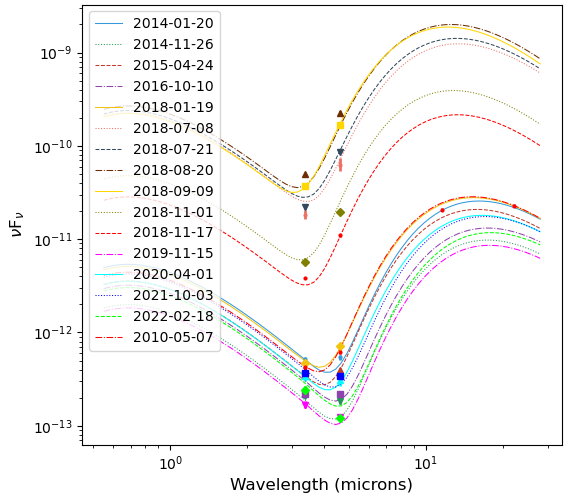} \quad
  \includegraphics[width=0.45\linewidth]{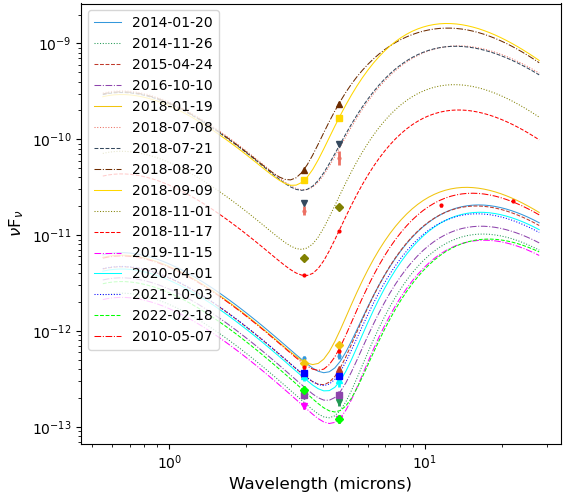} \\
  \includegraphics[width=0.44\linewidth, height=0.42\linewidth]{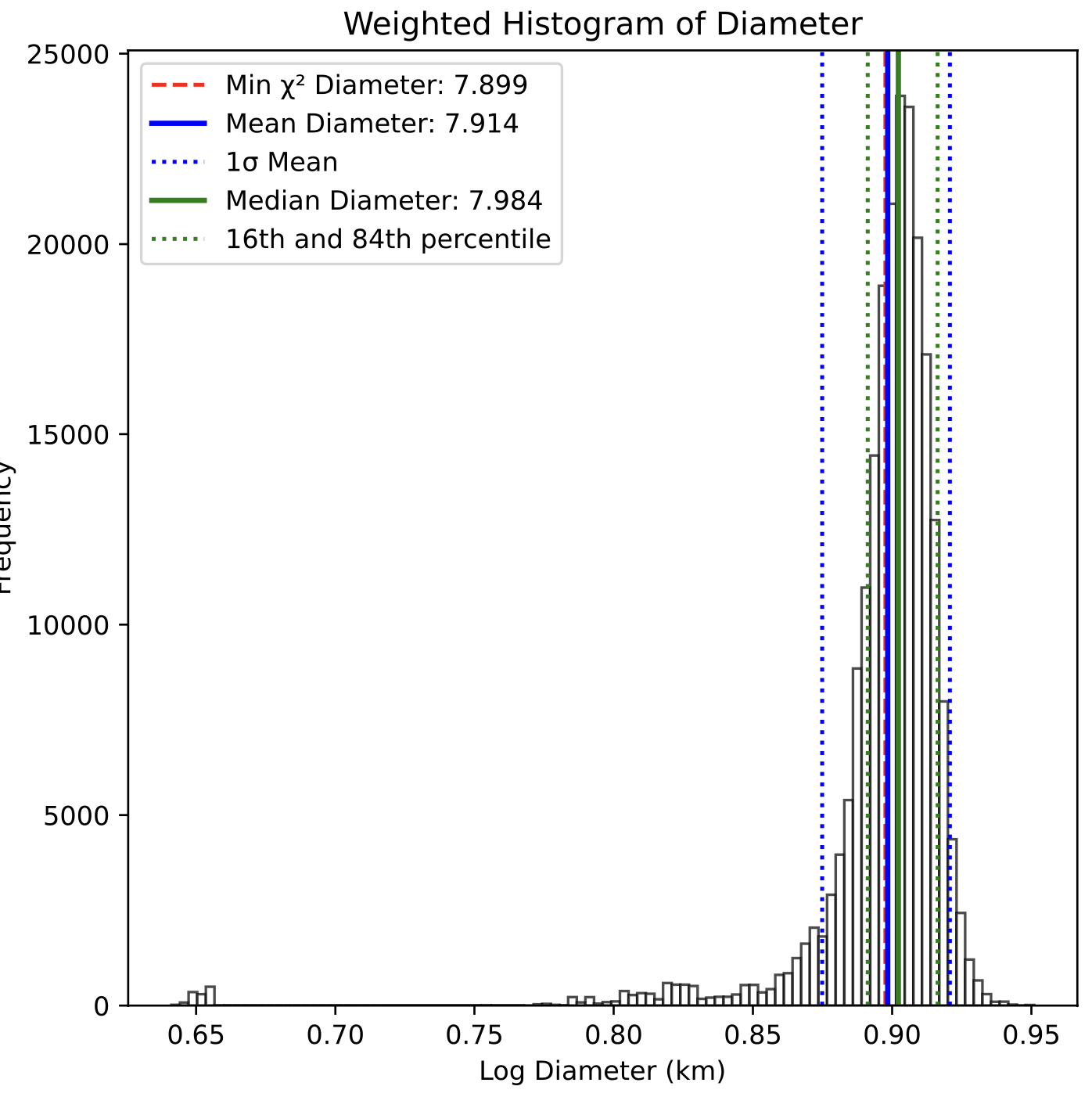} \quad
  \includegraphics[width=0.44\linewidth, height=0.42\linewidth]{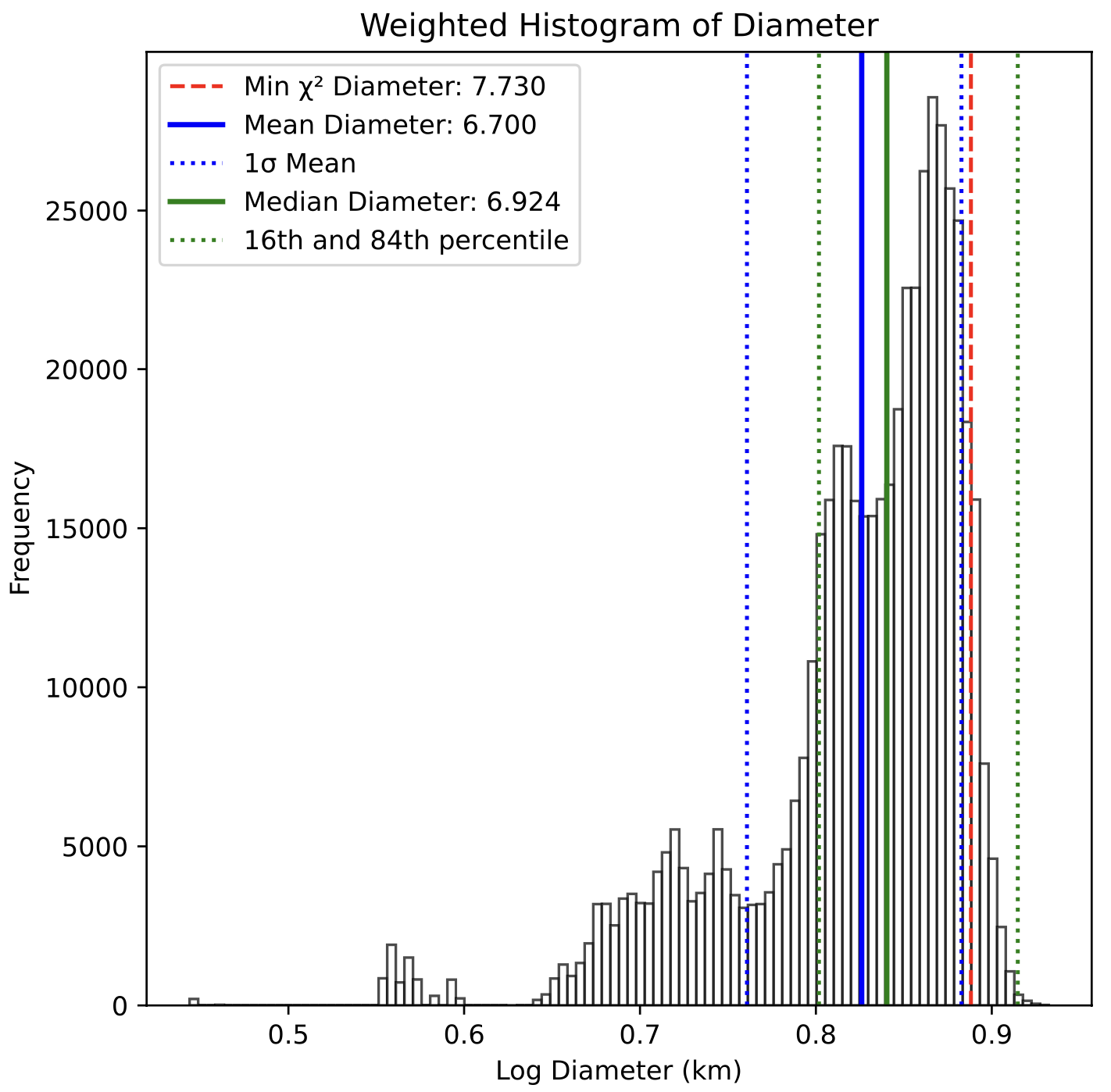}\\
  \includegraphics[width=0.45\linewidth, height=0.42\linewidth]{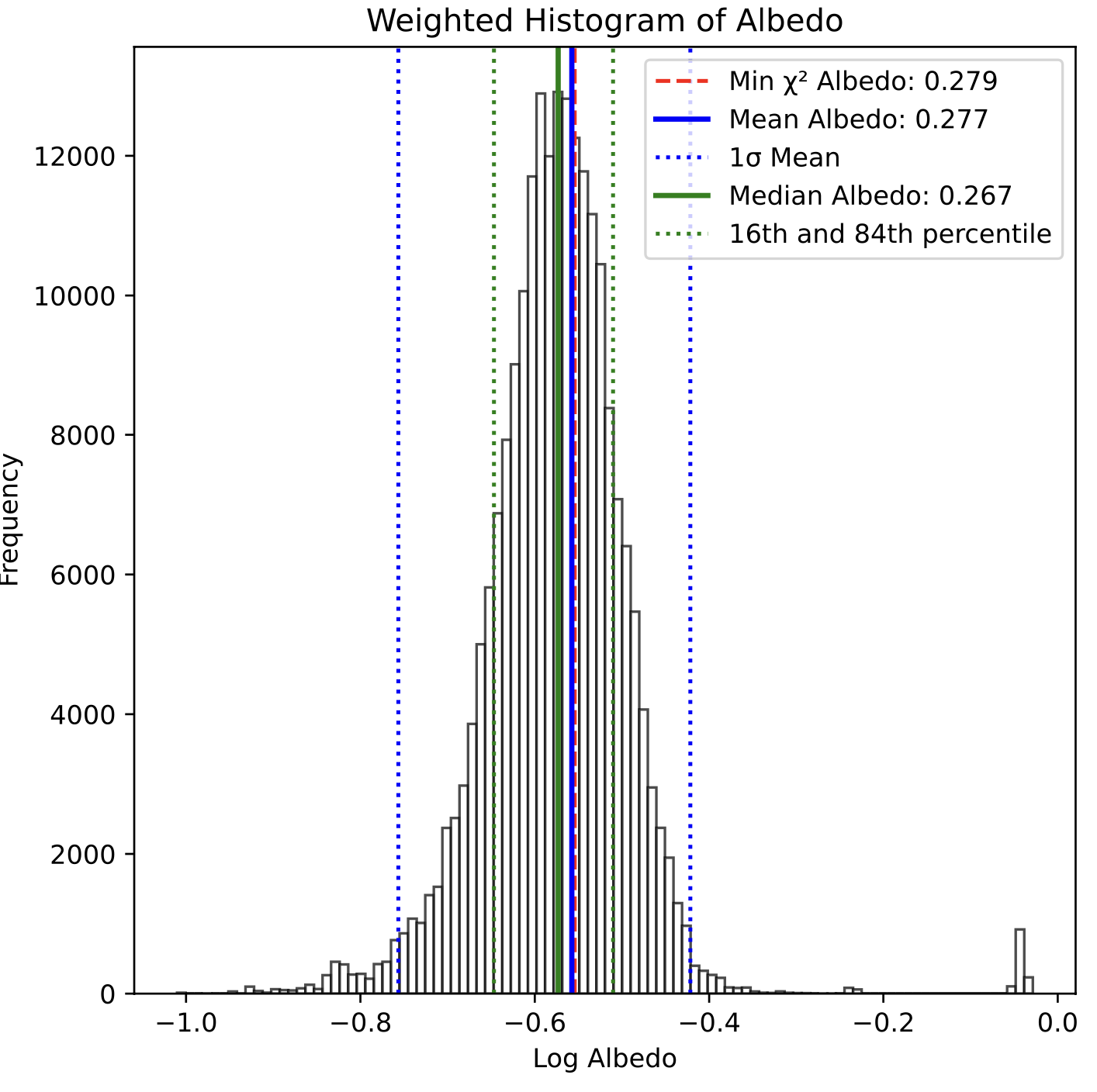} \quad
  \includegraphics[width=0.45\linewidth, height=0.42\linewidth]{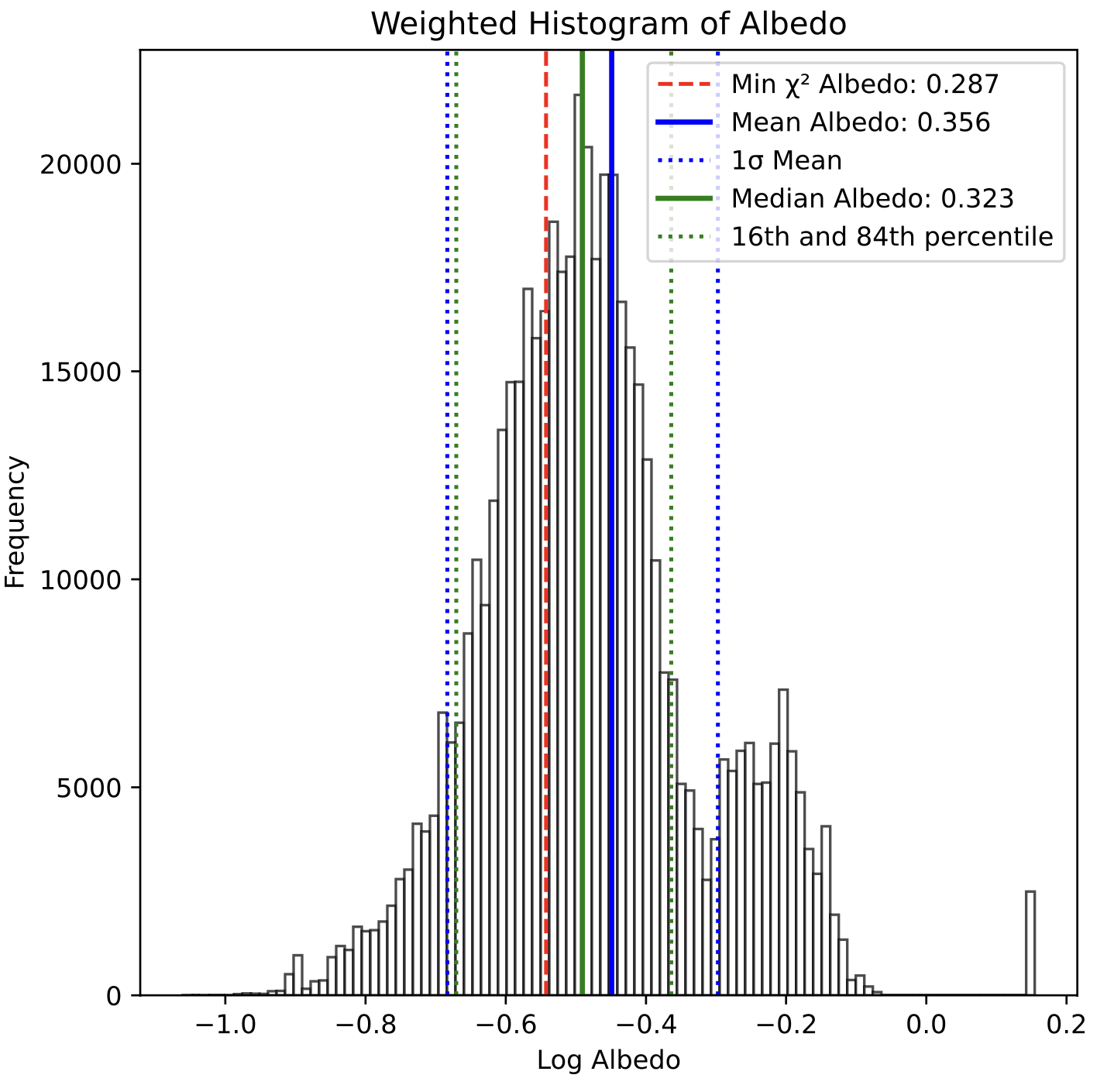}
  \caption{Upper: Spectral Energy Density fits for the spherical (left) and triaxial (right) model for the 16 epochs listed in Table \ref{tab:table1}. The normalization on the units of the y-axis allows for a more direct comparison between different wavelength regimes, as per \citet{wright07}. Lower: Diameter and albedo fits for the spherical (left) and triaxial (right) mdodel. The red dashed lines correspond to the single fits with the lowest $\chi^2$, while we show in solid blue and green, respectively, the weighted mean and median fits, with dotted lines delimiting their corresponding $1\sigma$ regions.}
  \label{fig:fig2}
\end{figure*}

    \subsection{Data and Methodology}
    As explained in Section \ref{sec.tpm}, thermophysical modelling is particularly effective if a large amount of data is available. Hence, for each target we queried the Infrared Science Archive to find all data available from all phases of the NEOWISE mission \citep{irsa}. The Catalog Search calculates the position of the target at a specific point in time and checks whether the telescope captured some emission there. This, however, can be entirely or in part due to background sources, polluting the detected flux. To reduce the possibility of this, the Image Search tool was used to check every detection returned by the Catalog and filter the data, in line with the process detailed in \citet{hung22}. NEOs emit brightly in the infrared, especially around the W3 channel wavelength, so most of the time detections from the cryo phase  were confirmed. The likelihood of background contamination increased for the post-cryo phase detections, where sometimes the detected source turned out to be brighter in W1 than W2, indicating a temperature consistent with a star instead of an asteroid.
At the end of this process, 16 epochs of observation (Table \ref{tab:table1}) were deemed acceptable and were then downloaded and prepared to be input into the thermophysical model. When an epoch contained several observations (due to the viewing geometry at the time), robust averaging was performed to reduce the impact of outliers. This is a technique consisting of a linear extrapolation-correction of $>2\sigma$ outlier points to reduce their impact on the mean. Also used were the best knowledge of the object's rotation period and absolute magnitude (see Section \ref{sec:resultstpm} for a discussion of the latter in relation to Section \ref{sec.hmag}). The TPM performs MCMC simulations to explore the parameter space and find the best fits to the provided data. Despite its heavy computational requirements, this TPM provides the ability to fit all epochs simultaneously, due to the use of physical parameters like thermal inertia.  We performed two fits: one assuming a spherical shape and  one that used a simple 3D ellipsoid \citep{wright07}, called ``triaxial" model. The latter requires a large number of epochs to fit the larger number of free parameters and takes a significantly longer time to run, but is a much more accurate description of the object, that is known to be non-spherical (see Section\ref{sec.ivar_intro}). Since this works aims to demonstrate the general validity of our approach, so that it will be possible to apply it to a larger population of asteroids in the future, we chose not to consider the available radar shape model for Ivar. In future studies, not all NEOs will indeed have an available radar shape model, but if enough infrared data was available for them, the triaxial model would be employable.

    \subsection{Results}
    \label{sec:resultstpm}
   The general results for the two models, obtained with $H_V = 12.43\pm0.02$, can be seen in Table \ref{tab:table2}. 
   The fitted diameters for both models agree quite well with previous results. The spherical model falls in same range as the result of 7.7 $\pm$0.6 km obtained by \citet{hanus15}, while the triaxial model gives axis ratios of $b/a = 0.334^{+0.086}_{-0.073}$ and $c/b = 0.883^{+0.094}_{-0.115}$. These naturally work both as semi-major axes' and maximum elongations' ratios. Therefore, we can combine them with the best-fit equivalent diameter from this model ($D=6.700\pm0.935$) using the formula from footnote \ref{D_formula}, giving as maximum elongations A = $14.507^{+0.325}_{-0.299}$ km, B = $4.845^{+1.252}_{-1.604}$ km, C = $4.278^{+1.196}_{-1.522}$ km. These are all in agreement with the elongations found by \citet{crowell17} and the prediction of ``maximum axis of minimum 7 km and probably within $20\%$ of 12 km" by \citet{ostro90} (Section\ref{sec.ivar_intro}). 

   In Figure \ref{fig:fig2} the improved fit obtained in the triaxial case for the target's SED across all observational epochs (error bars corresponding to the size of the dots) is apparent. This propagates to all fits and suggests a general greater reliability of the results from the triaxial model. An exception to this is the crater fraction, where the central result for the spherical model ($0.182^{+0.603}_{-0.144}$) falls in range with the one found by \citet{jones18}, where $f_c < 0.3$, better than the one obtained with the triaxial model. Both of our models, however, push the upper boundary of the crater function to values greater than 0.78.
   
\begin{table*}[b!]
    \centering
    \begin{tabular}{|c|c|c|c|c|c|c|}
    \hline
    Date & $V$ (mag)  & $\Delta$ (au) & $r$ (au)   & $\alpha$ ($^{\circ}$) & $P_{\rm r}$ (\%) & Observatory\\
    \hline
    2023-Apr-15 & 13.8 & 0.759 & 1.722 & 14.2 & $-0.68\pm0.09 $ & Calern\\
    2023-Apr-19 & 13.8 & 0.751 & 1.700 & 16.7 & $-0.23\pm0.08 $ & Calern\\
    2023-May-02 & 13.9 & 0.745 & 1.631 & 25.2 & $0.39\pm0.08 $ & Calern\\
    2024-Jan-13 & 15.9 & 1.707 & 1.616 & 34.3 & $1.50\pm0.10 $ & Rozhen\\
    \hline
    \end{tabular}
    \caption{Night average summary of Ivar polarization measurements ($P_{\rm r}$). $\Delta$ and $r$ corresponds to the distances of Ivar to Earth and Sun respectively. $\alpha$ is the solar phase angle. All parameters are listed for the mid-time of all observations. }
    \label{tab:Pola_obs}
    \end{table*}
    \hfill \newline
   In the bottom half of Figure \ref{fig:fig2}, we present weighted histograms of the fits explored by the MCMC routine for diameter (second row) and albedo (third row); we omit those for the other output parameters as their comparison to previous literature results are beyond the scope of this work. In the upper left, the fits for the diameter in the spherical model are tightly packed in a nice Gaussian-shape where least $\chi^2$ fit, mean and median are very close. This is linked to the duration of observation per each epoch ($\approx36$ hours), which means that enough different rotation states are sampled, and a good median is obtained \citep{masiero18}. In the triaxial case, the greater number of free parameters (such as the pole orientation), brings about the triple peak. This feature smooths out in the albedo-related diagrams, likely because the changes explored in the diameter phase space result in smaller shifts than the overall albedo uncertainty \citep{masiero21b}. 
   
   The albedo fits found, in both cases much higher than the previous literature values discussed in Section \ref{sec.comparison}, are consistent with each other by a $22\%$ margin when considering the central values and are well within each others' errors, thanks to the greater uncertainty on the triaxial model fits. Given the dependence of the TPM on the input absolute magnitude, we also run it using the value reported by the MPC ($H_V = 12.84$). We found that the diameter best fits for both the spherical and the triaxial case differed by $0.05\sigma$ from the ones presented in Table \ref{tab:table2}; the albedos can be thus computed directly, giving a range of 0.22-0.28 for the spherical model and 0.30-0.36 for the triaxial one. The consistency of these with the results obtained from polarimetry data is discussed in Section \ref{sec.comparison}.

  \section{Polarimetry}
    \label{sec.pol}
Polarimetric observations of solar system objects consist of measuring the linear degree of polarization of the unpolarized sunlight scattered by the surface of an atmosphere-less body. Upon scattering, the previously unpolarized light will be in a linear degree of polarization oriented either in the scattering plane (plane containing the Sun, the object, and the observer) or in the plane perpendicular to the scattering plane. For asteroids, we define the linear degree of polarization ($P_{\rm r}$) as the difference between the observed intensity of the light whose polarization is found to be in the plane perpendicular to the scattering plane and the intensity of the light whose polarization is found in the scattering plane, divided by the sum of these two intensity for normalization purposes. Thus, the $P_{\rm r}$ is defined as:

\begin{equation}
    P_{\rm r} = \dfrac{I_{\perp} - I_{\parallel}}{I_{\perp} + I_{\parallel}},
\end{equation}
where $I_{\perp}$ and $I_{\parallel}$ are respectively the intensity of the light whose polarization is perpendicular and in the scattering plane.

For all asteroids, $P_{\rm r}$ is displaying variation as a function of the phase angle, $\alpha$ (angle between the Sun, the object, and the observer), as per \citet{belskaya2015asteroid}. This dependence, called the phase-polarization curve, always displays the same characteristics for all asteroids. First the polarization is null at opposition ($\alpha$ = $0^{\circ}$), then $P_{\rm r}$ will be displaying negative values (polarization in the scattering plane) up to an inversion angle usually found around $\alpha = 20^{\circ}$. At larger phase angles the linear degree of polarization will rise linearly up to a maximum value, rarely observed, found around $80-120^{\circ}$. 
\begin{figure*}[t!]
  \centering
  \includegraphics[width=0.45\linewidth, height=0.4\linewidth]{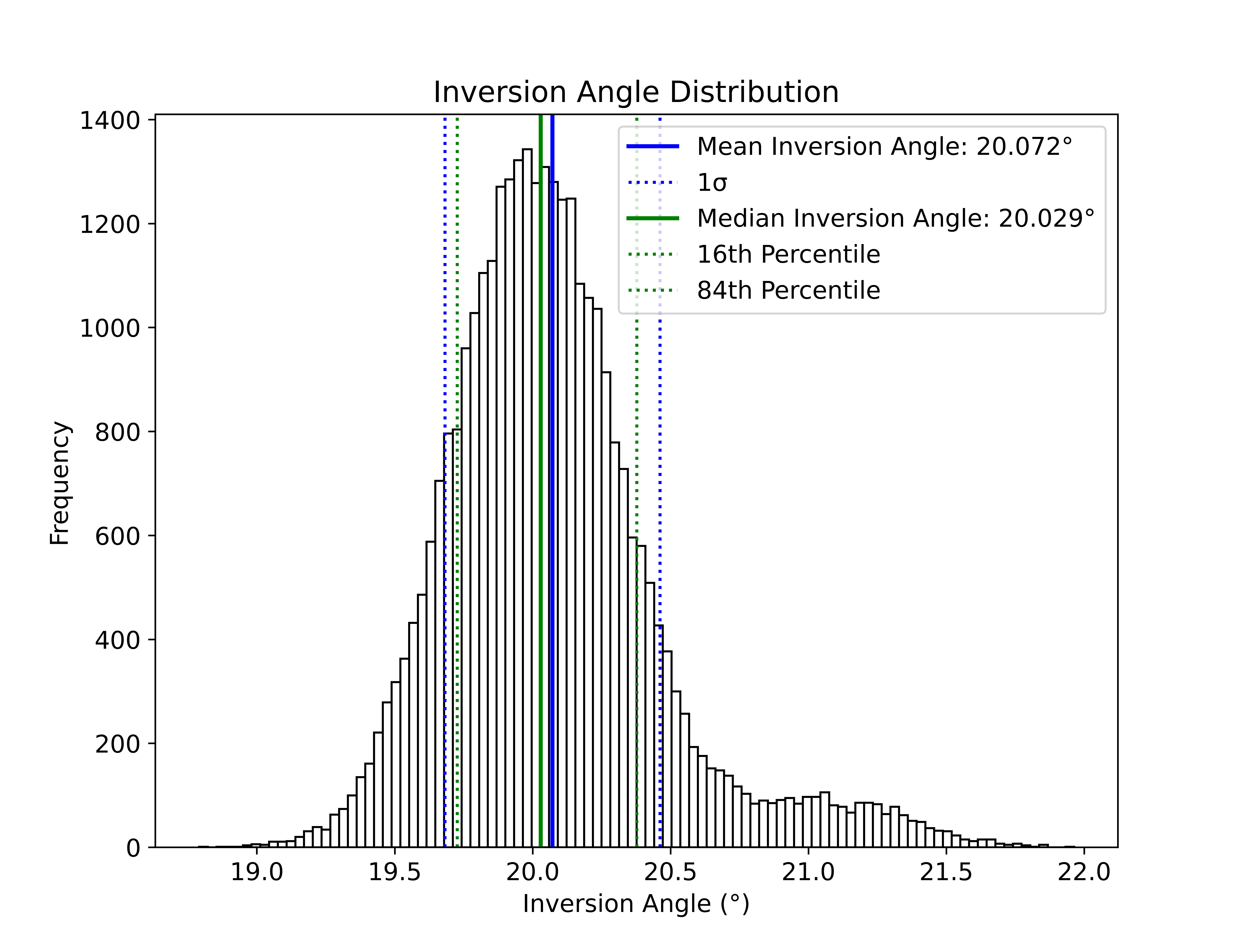} \quad
  \includegraphics[width=0.45\linewidth, height=0.4\linewidth]{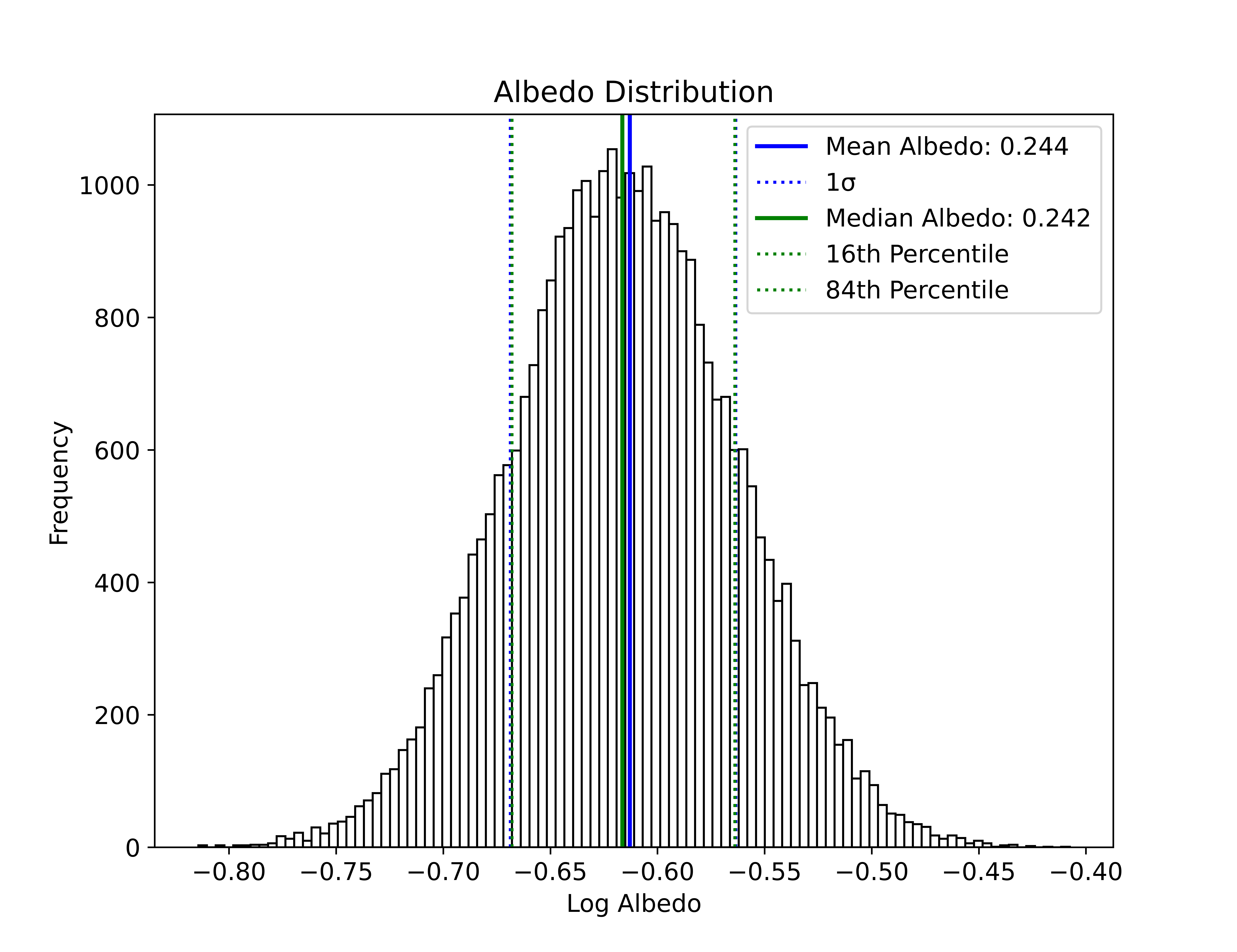} \\
  \caption{Fits for the inversion angle (left) and the albedo distribution (right) as explored by the MCMC routine introduced in \citet{devogele24}. }
  \label{fig:polmcmc}
\end{figure*}
 The absolute value of the linear polarization has been found to be linked to the albedo of the scattering surface, as per \citet{Umow_1905} and, more recently, by \citet{cellino15}. The brighter an asteroid is (high albedo) the lower the linear degree of polarization will be. This relation between albedo and polarization is called the Umov law \citep{Umow_1905} and has been used to determine albedo maps of the Moon using maximum value of polarization observed between $80-120^{\circ}$ \citep{Shkuratov_1992}. 

Unfortunately, asteroids can be rarely observed at such high phase angles. Proxies to the Umov law at lower phase angle have to be used to determine the albedo using polarimetric observations. The most used relation, up to now, is the \textit{slope-albedo} relation. This relation links the slope at inversion angle (i.e. when  $P_{\rm r}$ turns from negative to positive values around $\alpha = 20^{\circ}$) to the albedo. For a review of this relation, other polarimetric relations and the latest calibration of these see \citet{cellino15}. 
\subsection{Measurements and results}
We observed Ivar in polarimetry with the Torino Polarimeter (ToPol) \citep{devogele17} and the polarimetric mode of the two-channel focal reducer FoReRo2 \citep{jockers00}. ToPol is mounted on the Omicron-West 1.04m telescope from C2PU (Centre Pédagogique Planète et Univers) located at the Calern Observatory (MPC 010) near the city of Grasse (France). FoReRo2 is mounted on the 2m RCC telescope from the Bulgarian National Astronomical Observatory - Rozhen (MPC 071). 

With ToPol we obtained 14 observations with phase angles ranging from $4.24^{\circ}$ to $29.4^{\circ}$, to sample both the negative polarization branch (when $P_{\rm r}$ is negative) and the positive polarization branch (when $P_{\rm r}$ is positive) allowing to determine the slope at inversion angle. Eleven of these measurements have been obtained before 2022 and have been presented in Table B.1 of \citet{bendjoya22}. The three other measurements have been obtained in 2023. With FoReRo2 we obtained one polarimetric observations at a phase angle of $\alpha = 34.29^{\circ}$, allowing to further probe the positive polarization branch. The new polarimetric measurements presented in this paper are summarized in Table \ref{tab:Pola_obs}.
\begin{figure*}[t!]
  \centering
  \includegraphics[width=0.45\linewidth, height = 0.365\linewidth]{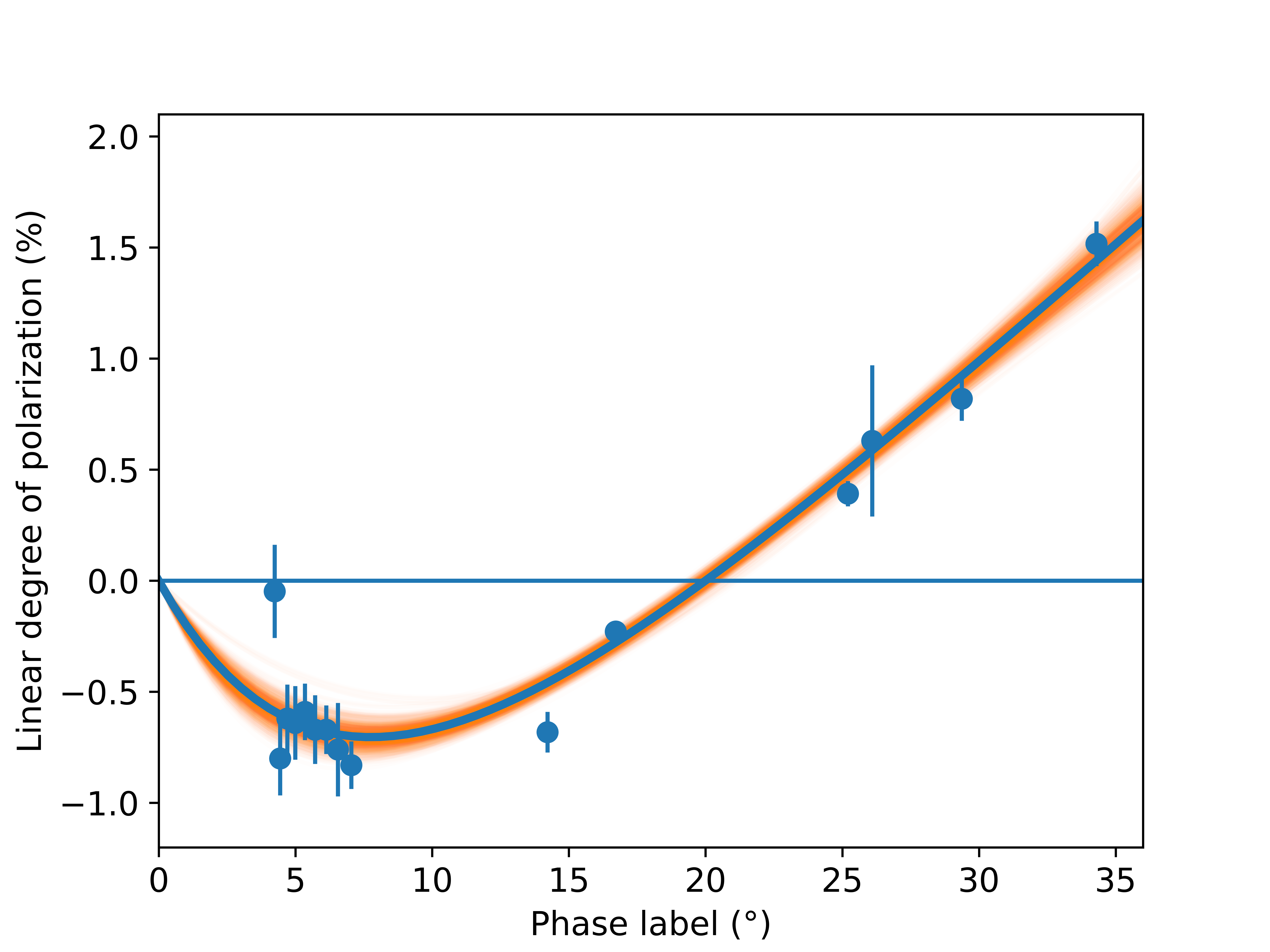} \quad
  \includegraphics[width=0.45\linewidth]{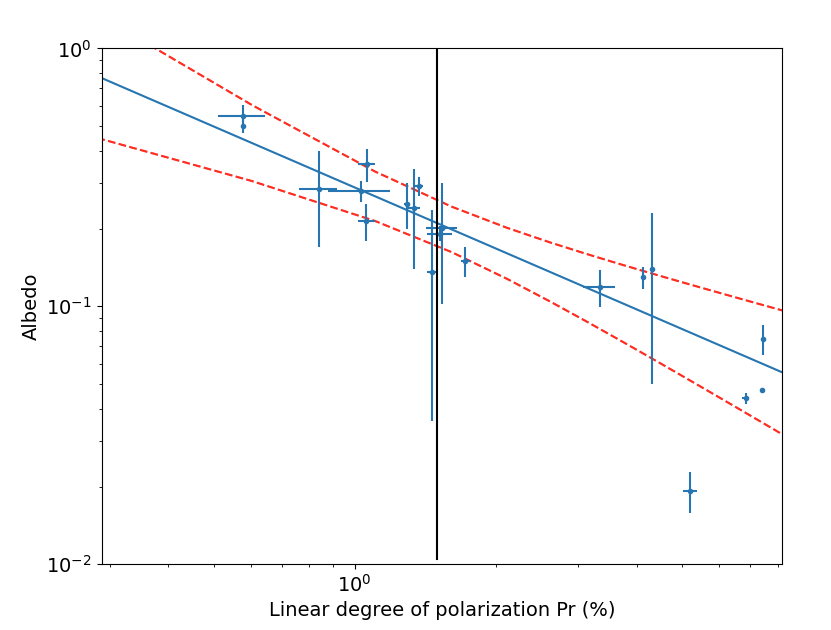} \\
  \caption{Left: Phase polarization curve of Ivar. The orange lines correspond to the fits explored by the MCMC before finding the best one (blue line). The 0 point at $4^{\circ}$ was ignored, which didn't affect the final results. Right: Resulting calibration at angle $\alpha = 34.29^{\circ}$. The asteroids used for the calibrations are: 1990 UQ, 1998 OR2 \citep{devogele24}, 1998 WT24 \citep{kiselev2002}, 1999 RM45, 2000 PQ9, 2000 SL, 2001 SN263, 2003 QQ47, 2005 UD \citep{Ishiguro_2022, Devogele_2020}, Apollo, Eros, Midas \citep[all three from][]{Bendjoya_2022}, Bennu \citep{cellino2018unusual}, Don Quixote \citep{Geem_2022}, Didymos \citep{gray2024polarimetry}, Hephaistos, Nereus, Phaethon \citep{kiselev2022asteroid, Okazaki_2020, Devogele_2018}, and Ra Shalom \citep{kiselev1999polarimetry}.}
  \label{fig:fig56}
\end{figure*}
We modeled the phase-polarization curve of Ivar using the exponential-linear \citet{muinonen09} relationship: 
\begin{equation}
    P_{\rm r}(\alpha) = A\left(\exp{\left(-\alpha/B\right)}-1\right)+C\alpha
\end{equation}
where $A$, $B$, and $C$ are parameters obtained through an optimizer. 
We explored the phase space using an MCMC fitting routine \citep[we used the same fitting routine and procedure as the one introduced and explained in][]{devogele24}, which provides a slope at inversion angle of $h = 0.088 \pm 0.003 \%/^{\circ}$ (see Fig. \ref{fig:polmcmc}, left, for the distribution of inversion angles explored). The prior on the inversion angle was generated following the lines of \citet{cellino2016polarimetric}. Using the \textit{slope-albedo} relation from \citep{cellino15}, we obtain an albedo of $p_{\rm V} = 0.24^{+0.04}_{-0.02}$.

In Fig. \ref{fig:fig56} (left), the phase-polarization curve of Ivar is displayed: the blue line represents the best fit of the exponential-linear model. Note that the $~0$ point around $4^{\circ}$ was ignored in plotting the best-fit: this didn't have any effect on the final result.

Using the highest phase angle measurements at $\alpha = 34.29^{\circ}$ with a linear degree of polarization $P_{\rm r} = 1.5\%$, we obtain a similar albedo value of $p_{\rm V} = 0.21\pm0.05$. This albedo is determined by calibrating the albedo-polarization relation at this specific phase angle, by comparing it with other asteroids for which polarization measurements at the same phase angle are available and for which albedo has been obtained through methods independent of polarimetry. As for the \textit{slope-albedo} relation, the value of linear polarization at a specific phase angle follows a linear dependence with the albedo of the asteroid in a log-log scale plot. Fig. \ref{fig:fig56} (right) shows our calibration at $\alpha = 34.29^{\circ}$. Each blue dots represents an asteroid for which polarization and albedo information are available, the blue line represents the best linear fit to the data with the dashed red lines the uncertainties on the fit. The black vertical line represents the polarization of Ivar at $\alpha = 34.29^{\circ}$: the intercept of this line with the blue one provides the albedo of Ivar according to this calibration.

  \section{Discussion}
  \label{sec.comparison}
The geometric albedo for 1672 Ivar is found, in all three cases, to be significantly higher than the one reported by the Small-Body Database Lookup of 0.15 and by several previous studies, such as 0.151 \citep{mainzer11}, $0.174\pm0.023$ \citep{mainzer14}, 0.15 \citep[NEATM from][considered ``the most reliable result judging from the thermal model fits"]{delbo03}. These  used the literature $H_V$ absolute magnitude; as this was significantly fainter than our value derived here, albedos calculated using the previous value would be expected to be shifted lower. Several indications of a possible higher value for Ivar's albedo can be however found in the literature: \citet{delbo03} itself obtained 0.20 using an STM, \citet{fornasier06} found a value of 0.3 using polarimetry, \citet{hanus15} fitted a varied shape thermal model to WISE data (VS-TPM) and obtained $0.255^{+0.02}_{-0.014}$. Our results (Table \ref{tab:albedo_final}, second column) follow this trend, and give a weighted mean of $0.26\pm0.02$. The value farthest from this is the result from the triaxial model fitting, which is not surprising since this model is significantly less well constrained than the spherical one. Similar considerations can be made for the result from Section \ref{sec.hmag}, where the uncertainty presented for the absolute magnitude is the error on the measurements - this is smaller than the actual dispersion, which is mostly due to lightcurve and viewing aspect variation. Moreover, the albedo result from this Section is dependent on the value of the diameter chosen for the calculation, which adds one further degree of freedom to the uncertainty. In the third column of Table \ref{tab:albedo_final}, we show for reference the values for albedo that we obtain in Section \ref{sec.hmag} and \ref{sec.tpm} when using the $H_V$ value reported by the MPC (all other literature values reported in Section \ref{sec.hmag} fall between this and 12.43, and would hence give albedos falling between the two columns' reported results). While they also seem to indicate a higher value for Ivar's albedo than 0.15, we note how much of an impact changing the input absolute magnitude has on them. As detailed in Section \ref{sec.hmag}, measurements of $H_V$ are varied and heavily dependent on the viewing geometry. It therefore becomes apparent how polarimetric observations, being independent of lightcurve or aspect effects, as well as much being less computationally-requiring, provide very reliable results, in line with other methods but unaffected by the variability introduced by absolute magnitude measurements. In light of this, we present $0.24^{+0.04}_{-0.02}$ as our proposed value for Ivar's albedo. Future work will aim to expand the catalogue of objects studied with these three methods, to further establish polarimetry as a valuable tool for the asteroids' characterization community.
  \begin{table*}[t!]
    \centering
    \begin{tabular}{|c|c|c|}
    \hline
    Technique & Albedo ($H_V = 12.43\pm0.02$) & Albedo (MPC: $H_V =12.83$) \\
    \hline
    Photometric measurements & $0.29\pm0.03$ & $0.20\pm0.02$\\
    Spherical TPM & $0.28\pm0.1$ & $0.22\pm0.04$\\
    Triaxial TPM  & $0.36\pm0.15$ & $0.30\pm0.1$\\
    Polarimetry - fitted slope at inversion angle  & $0.24^{+0.04}_{-0.02}$ & Unchanged \\
    Polarimetry - highest phase angle measurement  & $0.21\pm0.05$ & Unchanged\\
    \hline
    \end{tabular}
    \caption{Fits for the geometric albedo of Ivar (1627) obtained with the three techniques presented in this paper. In the second column, these are derived using an absolute magnitude of $H_V = 12.43\pm0.01$, obtained in Section \ref{sec.hmag}. In the third, we used the value reported by the MPC. All other values reported in Section \ref{sec.hmag} fall between this minimum and maximum: running the TPM would reflect this range, giving $0.22<p_V<0.28$ in the Spherical case, $0.30<p_V<0.36$ in the Triaxial one. Note instead how the results obtained with polarimetry are independent of the object's absolute magnitdue and size, so they are unaffected by the change of $H_V$ value in the two columns.}
    \label{tab:albedo_final}
    \end{table*}
  \section{Conclusions}
  In this work, we used three different techniques to constrain the physical parameters of the Amor class NEO Ivar (1627). Refined photometric measurements allowed us to find a brighter $H_V$ magnitude than the one recorded by the MPC, which was included in a thermophysical model fitting for the asteroid's diameter, albedo, thermal inertia and crater fraction, applied to both a spherical and ellipsoid shape. All the TPM fits were found to be consistent with the literature values, with the special case of the albedo result, which although higher than some of the commonly presented values around 0.15, is indeed consistent with previous literature works that calculated it to be $>0.20$. This result is also confirmed by the values obtained independently using both the previously mentioned photometric measurements and polarimetric data. Considerations on the impact that the highly-varying absolute magnitude value has on the results from the TPM and on direct derivations of the albedo (which are also dependent on size) lead us to identify polarimetry as a particularly efficient and accurate method. It only requires polarization measurements, that are independent on lightcurve or aspect effects, to derive a result that is validated both by literature and by the other two techniques. We therefore propose $0.24^{+0.04}_{-0.02}$ as a new value for Ivar's albedo. Work is undergoing to expand the catalog of objects for which the cross-reference approach presented here is undertaken, validating further the polarimetry method and establishing it as a streamlined tool to quickly derive Near Earth Asteroids' albedos and therefore their taxonomy classification.

  \section*{Acknowledgments}
  The authors acknowledge support for this work from the Yearly Opportunities for Planetary Defense Program under NASA grant 80NSSC22K0237.
  
  G.B. acknowledges partial support by grants: BG-175467353-2024-11-0020/KP-06-PH88/2 “Physical properties and chemical composition of asteroids and comets - a key to increasing our knowledge of the Solar System origin and evolution.” by the Bulgarian National Science Fund. All authors gratefully acknowledge observing grant support from the Institute of Astronomy and National Astronomical Observatory, Bulgarian Academy of Sciences.
  
  We thank the reviewers for the helpful comments and Dr. M. Mahlke and Dr. B. Carry for the fruitful discussion.
\footnotesize{
\bibliographystyle{aasjournal}
\bibliography{adssample.bib}}

\begin{thebibliography}{}
\expandafter\ifx\csname natexlab\endcsname\relax\def\natexlab#1{#1}\fi
\providecommand{\url}[1]{\href{#1}{#1}}
\providecommand{\dodoi}[1]{doi:~\href{http://doi.org/#1}{\nolinkurl{#1}}}
\providecommand{\doeprint}[1]{\href{http://ascl.net/#1}{\nolinkurl{http://ascl.net/#1}}}
\providecommand{\doarXiv}[1]{\href{https://arxiv.org/abs/#1}{\nolinkurl{https://arxiv.org/abs/#1}}}

\bibitem[{Belskaya {et~al.}(2015)Belskaya, Cellino, Gil-Hutton, Muinonen,
  Shkuratov, {et~al.}}]{belskaya2015asteroid}
Belskaya, I., Cellino, A., Gil-Hutton, R., {et~al.} 2015, Asteroids iv, 151

\bibitem[{Bendjoya {et~al.}(2022{\natexlab{a}})Bendjoya, Cellino, Rivet,
  Devog{\`e}le, Bagnulo, Abe, Vernet, Gil-Hutton, \& Veneziani}]{bendjoya22}
Bendjoya, P., Cellino, A., Rivet, J.-P., {et~al.} 2022{\natexlab{a}}, Astronomy
  \& Astrophysics, 665, A66

\bibitem[{Bendjoya {et~al.}(2022{\natexlab{b}})Bendjoya, Cellino, Rivet,
  Devog{\`e}le, Bagnulo, Abe, Vernet, Gil-Hutton, \& Veneziani}]{Bendjoya_2022}
---. 2022{\natexlab{b}}, Astronomy \& Astrophysics, 665, A66

\bibitem[{Binzel {et~al.}(2015)}]{binzel15}
Binzel, R.~P., {et~al.} 2015, Asteroids IV

\bibitem[{Carry {et~al.}(2024)Carry, Peloton, Le~Montagner, Mahlke, \&
  Berthier}]{carry24}
Carry, B., Peloton, J., Le~Montagner, R., Mahlke, M., \& Berthier, J. 2024,
  Astronomy and Astrophysics

\bibitem[{Carry {et~al.}(2016)Carry, Solano, Eggl, \& DeMeo}]{carry16}
Carry, B., Solano, E., Eggl, S., \& DeMeo, F.~E. 2016, Icarus

\bibitem[{Cellino {et~al.}(2018)Cellino, Bagnulo, Belskaya, \&
  Christou}]{cellino2018unusual}
Cellino, A., Bagnulo, S., Belskaya, I., \& Christou, A. 2018, Monthly Notices
  of the Royal Astronomical Society: Letters, 481, L49

\bibitem[{Cellino {et~al.}(2016)Cellino, Bagnulo, Gil-Hutton, Tanga,
  Canada-Assandri, \& Tedesco}]{cellino2016polarimetric}
Cellino, A., Bagnulo, S., Gil-Hutton, R., {et~al.} 2016, Monthly Notices of the
  Royal Astronomical Society, 455, 2091

\bibitem[{Cellino {et~al.}(2015)Cellino, Bagnulo, Gil-Hutton,
  {et~al.}}]{cellino15}
---. 2015, \mnras, 451, 3473, \dodoi{10.1093/mnras/stv1188}

\bibitem[{Crowell(2017)}]{crowell17}
Crowell, J. e.~a. 2017, Icarus, 291

\bibitem[{Delbo(2015)}]{delbo15}
Delbo, M. e.~a. 2015, Asteroids IV

\bibitem[{Delbó {et~al.}(2003)Delbó, Harris, Binzel, \& Davies}]{delbo03}
Delbó, M., Harris, A.~W., Binzel, Richard P.and~Pravec, P., \& Davies, J.~K.
  2003, Icarus, 166

\bibitem[{Devog{\`e}le {et~al.}(2024)Devog{\`e}le, McGilvray, MacLennan,
  {et~al.}}]{devogele24}
Devog{\`e}le, M., McGilvray, A., MacLennan, E., {et~al.} 2024, The Planetary
  Science Journal, 5, 44, \dodoi{10.3847/PSJ/ad1f70}

\bibitem[{Devog{\`e}le {et~al.}(2017)Devog{\`e}le, Cellino, Bagnulo, Rivet,
  Bendjoya, Abe, Pernechele, Massone, Vernet, Tanga, \& Dimur}]{devogele17}
Devog{\`e}le, M., Cellino, A., Bagnulo, S., {et~al.} 2017, Monthly Notices of
  the Royal Astronomical Society, 465, 4335, \dodoi{10.1093/mnras/stw2952}

\bibitem[{Devog{\`e}le {et~al.}(2018)Devog{\`e}le, Cellino, Borisov, Bendjoya,
  Rivet, Abe, Bagnulo, Christou, Vernet, Donchev, {et~al.}}]{Devogele_2018}
Devog{\`e}le, M., Cellino, A., Borisov, G., {et~al.} 2018, Monthly Notices of
  the Royal Astronomical Society, 479, 3498

\bibitem[{Devog{\`e}le {et~al.}(2020)Devog{\`e}le, MacLennan, Gustafsson,
  Moskovitz, Chatelain, Borisov, Abe, Arai, Fedorets, Ferrais,
  {et~al.}}]{Devogele_2020}
Devog{\`e}le, M., MacLennan, E., Gustafsson, A., {et~al.} 2020, The Planetary
  Science Journal, 1, 15

\bibitem[{Dotson {et~al.}(2024)Dotson, Wheeler, \& Mathias}]{Dotson_2024}
Dotson, J.~L., Wheeler, L., \& Mathias, D. 2024, Acta Astronautica

\bibitem[{Durech {et~al.}(2010)Durech, Sidorin, \& Kaasalainen}]{durech10}
Durech, J., Sidorin, V., \& Kaasalainen, M. 2010, Astronomy \& Astrophysics, 513

\bibitem[{Fenucci {et~al.}(2023)Fenucci, Novakovi{\'c}, \&
  Mar{\v{c}}eta}]{Fenucci_2023}
Fenucci, M., Novakovi{\'c}, B., \& Mar{\v{c}}eta, D. 2023, Astronomy \&
  Astrophysics, 675, A134

\bibitem[{Foreman-Mackey {et~al.}(2013)Foreman-Mackey, Hogg, Lang, \&
  Goodman}]{foreman13}
Foreman-Mackey, D., Hogg, D.~W., Lang, D., \& Goodman, J. 2013, The
  Astronomical Society of the Pacific

\bibitem[{Fornasier {et~al.}(2006)Fornasier, Belskaya, Shkuratov, Pernechele,
  Barbieri, Giro, \& Navasardyan}]{fornasier06}
Fornasier, S., Belskaya, I.~N., Shkuratov, Y.~G., {et~al.} 2006, Astronomy \&
  Astrophysics

\bibitem[{Geem {et~al.}(2022)Geem, Ishiguro, Bach, Kuroda, Naito, Hanayama,
  Kim, Kwon, Jin, Sekiguchi, {et~al.}}]{Geem_2022}
Geem, J., Ishiguro, M., Bach, Y.~P., {et~al.} 2022, Astronomy \& Astrophysics,
  658, A158

\bibitem[{Gray {et~al.}(2024)Gray, Bagnulo, Granvik, Cellino, Jones,
  Kolokolova, Moreno, Muinonen, Mu{\~n}oz, Opitom,
  {et~al.}}]{gray2024polarimetry}
Gray, Z., Bagnulo, S., Granvik, M., {et~al.} 2024, The Planetary Science
  Journal, 5, 18

\bibitem[{Hanuš {et~al.}(2015)Hanuš, Delbo’, Ďurech, \&
  Alí-Lagoa}]{hanus15}
Hanuš, J., Delbo’, M., Ďurech, J., \& Alí-Lagoa, V. 2015, Icarus, 256, 101

\bibitem[{Harris \& Young(1985)}]{harris85}
Harris, A.~W., \& Young, J.~W. 1985, Bulletin of the American Astronomical
  Society, 17

\bibitem[{Hoffmann {et~al.}(2024)Hoffmann, Micheli, Cano, Devog{\`e}le,
  Farnocchia, Pravec, Vere{\v{s}}, \& Poppe}]{Hoffmann_2024}
Hoffmann, T., Micheli, M., Cano, J.~L., {et~al.} 2024, arXiv preprint
  arXiv:2408.07474

\bibitem[{Hung {et~al.}(2022)Hung, Hanuš, Masiero, \& Tholen}]{hung22}
Hung, D., Hanuš, J., Masiero, J.~R., \& Tholen, D.~J. 2022, Planetary Science
  Journal

\bibitem[{Ishiguro {et~al.}(2022)Ishiguro, Bach, Geem, Naito, Kuroda, Im, Lee,
  Seo, Jin, Kwon, {et~al.}}]{Ishiguro_2022}
Ishiguro, M., Bach, Y.~P., Geem, J., {et~al.} 2022, Monthly Notices of the
  Royal Astronomical Society, 509, 4128

\bibitem[{Jockers {et~al.}(2000)Jockers, Credner, Bonev, \& et~al.}]{jockers00}
Jockers, K., Credner, T., Bonev, T., \& et~al. 2000, Kinematika i Fizika
  Nebesnykh Tel Supplement, 3, 13

\bibitem[{Jones(2018)}]{jones18}
Jones, J. 2018, Electronic Theses and Dissertations, 5838

\bibitem[{Kaasalainen {et~al.}(2004)}]{kaasalainen04}
Kaasalainen, M., {et~al.} 2004, Icarus, 167, 178

\bibitem[{Kiselev {et~al.}(1999)Kiselev, Rosenbush, \&
  Jockers}]{kiselev1999polarimetry}
Kiselev, N., Rosenbush, V., \& Jockers, K. 1999, Icarus, 140, 464

\bibitem[{Kiselev {et~al.}(2002)Kiselev, Rosenbush, Jockers, Velichko,
  Shakhovskoj, Efimov, Lupishko, \& Rumyantsev}]{kiselev2002}
Kiselev, N., Rosenbush, V., Jockers, K., {et~al.} 2002, in Asteroids, Comets,
  and Meteors: ACM 2002, Vol. 500, 887--890

\bibitem[{Kiselev {et~al.}(2022)Kiselev, Rosenbush, Petrov, Luk'yanyk, Ivanova,
  Pit, Antoniuk, \& Afanasiev}]{kiselev2022asteroid}
Kiselev, N., Rosenbush, V., Petrov, D., {et~al.} 2022, Monthly Notices of the
  Royal Astronomical Society, 514, 4861

\bibitem[{Koren {et~al.}(2015)Koren, Wright, \& Mainzer}]{koren15}
Koren, S.~C., Wright, E.~L., \& Mainzer, A. 2015, Icarus, 258, 82

\bibitem[{Kueny {et~al.}(2023)Kueny, Chandler, Devog{\'e}le, Moskovitz, Pravec,
  Ku{\v{c}}{\'a}kov{\'a}, Hornoch, Ku{\v{s}}nir{\'a}k, Granvik,
  Konstantopoulou, {et~al.}}]{Kueny_2023}
Kueny, J.~K., Chandler, C.~O., Devog{\'e}le, M., {et~al.} 2023, The Planetary
  Science Journal, 4, 56

\bibitem[{Lauretta {et~al.}(2017)Lauretta, Balram-Knutson, Beshore, Boynton,
  Drouet~d'Aubigny, DellaGiustina, Enos, Golish, Hergenrother, Howell, Bennett,
  {et~al.}}]{lauretta17}
Lauretta, D.~S., Balram-Knutson, S.~S., Beshore, E., {et~al.} 2017, Space
  Science Reviews, 212, 925

\bibitem[{Magri {et~al.}(2011)Magri, Howell, Nolan, Taylor, Fern{\'a}ndez,
  Mueller, Vervack~Jr, Benner, Giorgini, Ostro, {et~al.}}]{Magri_2011}
Magri, C., Howell, E.~S., Nolan, M.~C., {et~al.} 2011, Icarus, 214, 210

\bibitem[{Mahlke {et~al.}(2021{\natexlab{a}})Mahlke, Carry, \&
  Denneau}]{Mahlke_2021}
Mahlke, M., Carry, B., \& Denneau, L. 2021{\natexlab{a}}, Icarus, 354, 114094

\bibitem[{Mahlke {et~al.}(2021{\natexlab{b}})Mahlke, Carry, \&
  Denneau}]{mahlke21}
---. 2021{\natexlab{b}}, Icarus

\bibitem[{Mainzer {et~al.}(2011a)Mainzer, Grav, Masiero, {et~al.}}]{mainzer11}
Mainzer, A., Grav, T., Masiero, J., {et~al.} 2011a, \apj, 736, 100,
  \dodoi{10.1088/0004-637X/736/2/100}

\bibitem[{Mainzer {et~al.}(2011b)Mainzer, Grav, Masiero,
  {et~al.}}]{mainzer11tax}
---. 2011b, \apj, 741, 90, \dodoi{10.1088/0004-637X/741/2/90}

\bibitem[{Mainzer {et~al.}(2014)}]{mainzer14}
Mainzer, A., {et~al.} 2014, ApJ, 792, 30

\bibitem[{Masci {et~al.}(2018)Masci, Laher, Rusholme, Shupe, Groom, Surace,
  Jackson, Monkewitz1, Beck, \& Flynn}]{masci18}
Masci, F.~J., Laher, R.~R., Rusholme, B., {et~al.} 2018, Publications of the
  Astronomical Society of the Pacific

\bibitem[{Masiero {et~al.}(2021{\natexlab{a}})Masiero, Mainzer, Bauer, Cutri,
  Grav, Kramer, Pittichov{\'a}, \& Wright}]{Masiero_2021}
Masiero, J.~R., Mainzer, A., Bauer, J., {et~al.} 2021{\natexlab{a}}, The
  Planetary Science Journal, 2, 162

\bibitem[{Masiero {et~al.}(2018)Masiero, Redwing, Mainzer,
  {et~al.}}]{masiero18}
Masiero, J.~R., Redwing, E., Mainzer, A.~K., {et~al.} 2018, \aj, 156, 60,
  \dodoi{10.3847/1538-3881/aacce4}

\bibitem[{Masiero {et~al.}(2019)Masiero, Wright, \& Mainzer}]{masiero19}
Masiero, J.~R., Wright, E.~L., \& Mainzer, A.~K. 2019, AJ, 158, 97,
  \dodoi{10.3847/1538-3881/ab31a6}

\bibitem[{Masiero {et~al.}(2021{\natexlab{b}})Masiero, Wright, \&
  Mainzer}]{masiero21b}
---. 2021{\natexlab{b}}, The Planetary Science J, 2, 32,
  \dodoi{10.3847/PSJ/abda4d}

\bibitem[{Masiero {et~al.}(2012)}]{masiero12}
Masiero, J.~R., {et~al.} 2012, ApJ, 749, 104

\bibitem[{Mueller {et~al.}(2011)Mueller, Delbo', Hora, Trilling, Bhattacharya,
  \& Bottke}]{mueller11}
Mueller, M., Delbo', M., Hora, J.~L., {et~al.} 2011, \aj

\bibitem[{Muinonen {et~al.}(2010)Muinonen, Belskaya, Cellino, Delb{\`o},
  Levasseur-Regourd, Penttil{\"a}, \& Tedesco}]{muinonen10}
Muinonen, K., Belskaya, I.~N., Cellino, A., {et~al.} 2010, Icarus, 209, 542

\bibitem[{Muinonen {et~al.}(2009)Muinonen, Penttilä, Cellino, Belskaya,
  Delbò, Levasseur-Regourd, \& Tedesco}]{muinonen09}
Muinonen, K., Penttilä, A., Cellino, A., {et~al.} 2009, Icarus 209, 542

\bibitem[{Okazaki {et~al.}(2020)Okazaki, Sekiguchi, Ishiguro, Naito, Urakawa,
  Imai, Ono, Warner, \& Watanabe}]{Okazaki_2020}
Okazaki, R., Sekiguchi, T., Ishiguro, M., {et~al.} 2020, Planetary and Space
  Science, 180, 104774

\bibitem[{Ostro {et~al.}(1990)Ostro, Campbell, Hine, Shapiro, Chandler, Werner,
  \& Rosema}]{ostro90}
Ostro, S.~J., Campbell, D.~B., Hine, A.~A., {et~al.} 1990, AJ, 99, 2012

\bibitem[{Rivkin {et~al.}(2021)}]{rivkin21}
Rivkin, A.~S., {et~al.} 2021, Planet. Sci. J., 2, 173

\bibitem[{Satpathy {et~al.}(2022)Satpathy, Mainzer, Masiero, Linder, Cutri,
  Wright, Pittichová, Grav, \& Kramer}]{satpathy22}
Satpathy, A., Mainzer, A., Masiero, J.~R., {et~al.} 2022, PSJ

\bibitem[{Shkuratov \& Opanasenko(1992)}]{Shkuratov_1992}
Shkuratov, Y.~G., \& Opanasenko, N. 1992, Icarus, 99, 468

\bibitem[{Thirouin {et~al.}(2016)Thirouin, Moskovitz, Binzel, Christensen,
  DeMeo, Person, Polishook, Thomas, Trilling, Willman,
  {et~al.}}]{Thirouin_2016}
Thirouin, A., Moskovitz, N., Binzel, R., {et~al.} 2016, The Astronomical
  Journal, 152, 163

\bibitem[{Tonry {et~al.}(2018)Tonry, Denneau, Heinze, Stalder, Smith, Smartt,
  Stubbs, Weiland, \& Rest}]{Tonry_2018}
Tonry, J., Denneau, L., Heinze, A., {et~al.} 2018, Publications of the
  Astronomical Society of the Pacific, 130, 064505

\bibitem[{Trilling {et~al.}(2008)Trilling, Mueller, Hora, Fazio, Spahr,
  Stansberry, Smith, Chesley, \& Mainzer}]{Trilling_2008}
Trilling, D.~E., Mueller, M., Hora, J., {et~al.} 2008, The Astrophysical
  Journal, 683, L199

\bibitem[{Umow(1905)}]{Umow_1905}
Umow, v.~N. 1905, Phys. Z, 6, 674

\bibitem[{Vega \& Devogele(2023)}]{Vega_2023}
Vega, N., \& Devogele, M. 2023, LPI Contributions, 2851, 2590

\bibitem[{Warner(2015)}]{warner15}
Warner, B.~D. 2015, Minor Planet Bull., 42, 172

\bibitem[{Watanabe {et~al.}(2017)Watanabe, Tsuda, Yoshikawa, Tanaka, Saiki, \&
  Nakazawa}]{watanabe17}
Watanabe, S., Tsuda, Y., Yoshikawa, M., {et~al.} 2017, Space Science Reviews,
  208, 3

\bibitem[{Wright(2007)}]{wright07}
Wright, E.~L. 2007, astro-ph/0703085, \dodoi{10.48550/arXiv.astro-ph/0703085}

\bibitem[{Wright {et~al.}(2016)Wright, Mainzer, Masiero, Grav, \&
  Bauer}]{wright16}
Wright, E.~L., Mainzer, A., Masiero, J., Grav, T., \& Bauer, J. 2016, AJ, 152

\bibitem[{Wright {et~al.}(2010)}]{wright10}
Wright, E.~L., {et~al.} 2010, AJ, 140, 1868

\bibitem[{Wright {et~al.}(2019)Wright, Eisenhardt, Mainzer, Ressler, Cutri,
  Jarrett, Kirkpatrick, Padgett, {et~al.}}]{irsa}
Wright, E.~L., Eisenhardt, P. R.~M., Mainzer, A.~K., {et~al.} 2019, IPAC,
  \dodoi{10.26131/IRSA1}

\bibitem[{Xie {et~al.}(2021)Xie, Bennett, \& Dempster}]{Xie_2021}
Xie, R., Bennett, N.~J., \& Dempster, A.~G. 2021, Acta Astronautica, 181, 249

\end{thebibliography}
    \end{multicols}

\end{document}